\documentclass{aa}
\usepackage{graphicx}

\begin{document}

\title{Physical parameters of the high-mass X-ray binary \object{4U1700-37}\thanks{Based on 
observations collected at the European Southern Observatory, La Silla, Chile (64.H-0224)}}
\author{J.~S.~Clark\inst{1} 
\and S.~P.~Goodwin\inst{2} 
\and P.~A.~Crowther\inst{1}
\and L.~Kaper\inst{3}
\and M.~Fairbairn\inst{4}
\and N.~Langer\inst{5} 
\and C.~Brocksopp\inst{6}
}

\offprints{J. S. Clark, jsc@star.ucl.ac.uk}

\institute{ Department of Physics and Astronomy, University College London,
Gower Street, London, WC1E 6BT, England, UK  
\and Department of Physics and Astronomy, University of Wales, Cardiff,
 CF24 3YB, Wales, UK.  
\and Astronomical Institute ``Anton Pannekoek'', University of Amsterdam and Center for High-Energy 
Astrophysics, Kruislaan 403, 1098 SJ Amsterdam, the Netherlands
\and Service de Physique Th\'eorique, CP225,
Universit\'e Libre de Bruxelles, B-1050 Brussels, Belgium
\and Astronomical Institute, Utrecht University, Princetonplein 5, NL-3584 CC, Utrecht, The Netherlands
\and Astrophysics Research Institute, Liverpool John Moores
University, Liverpool, L41 1LD, U.K.}  

\date{Received    / Accepted     }

\abstract{We present the results of a detailed non-LTE analysis of the
ultraviolet and optical spectrum of the O6.5\,Iaf$^+$ star
\object{HD~153919} - the mass donor in the high-mass X-ray binary
\object{4U1700-37}. We find that the star has a luminosity
log(L$_{\ast}$/L$_{\odot}$)=5.82$\pm$0.07, T$_{\rm
eff}$=35,000$\pm$1000~K, radius
R$_{\ast}$=21.9$^{+1.3}_{-0.5}$~R$_{\odot}$, mass-loss rate
\.{M}=9.5$\times$10$^{-6}$~M$_{\odot}$~yr$^{-1}$, and a significant
overabundance of nitrogen (and possibly carbon) relative to solar
values.  Given the eclipsing nature of the system these results allow
us to determine the most likely masses of both components of the
binary via Monte Carlo simulations. These suggest a mass for
\object{HD~153919} of M$_{\ast} = 58 \pm 11$~M$_{\odot}$ - implying the
initial mass of the companion was rather high
($\ga$60~M$_{\odot}$). The most likely mass for the compact companion
is found to be M$_x$ =2.44$\pm$0.27~M$_{\odot}$, with only 3.5~per
cent of the trials resulting in a mass less than 2.0~M$_{\odot}$ and
none less than 1.65~M$_{\odot}$. Such a value is significantly in
excess of the upper observational limit to the masses of neutron stars
of 1.45~M$_{\odot}$ found by Thorsett \& Chakrabarthy
(\cite{thorsett}), although a mass of 1.86~M$_{\odot}$ has recently
been reported for the \object{Vela X-1} pulsar (Barziv et
al. \cite{barziv}).  Our observational data is inconsistent with the
canonical neutron star mass and the lowest black hole mass observed
($\ga$4.4~M$_{\odot}$; \object{Nova Vel}).  Significantly changing
observational parameters {\em can} force the compact object mass into
either of these regimes but, given the strong proportionality between
M$_{\ast}$ and M$_x$, the O-star mass changes by factors of greater than 2,
well beyond the limits determined from its evolutionary state and
surface gravity.  The low mass of the compact object implies that it
is difficult to form high mass black holes through both the Case A \&
B mass transfer channels and, if the compact object is a neutron star,
would significantly constrain the high density nuclear equation of
state.  }

\titlerunning{Physical parameters of 4U~1700-37}

\maketitle

\section{Introduction}   

First detected by the {\em Uhuru} satellite (Jones et
al. \cite{jones}) the eclipsing X-ray source \object{4U~1700-37} was
quickly associated with the luminous O6.5 Iaf$^{+}$ star
\object{HD~153919}, confirming \object{4U~1700-37} as a high-mass
X-ray binary (henceforth HMXB). HMXBs are systems composed of an OB
star and a compact companion (neutron star or black hole), with the
X-ray emission resulting from the accretion of material by the compact
companion. In the subclass of supergiant HMXB systems material is
accreted either via Roche lobe overflow or directly from the powerful
stellar wind of the OB primary. Given that \object{HD~153919} slightly
underfills its Roche Lobe (e.g. Conti \cite{conti}) mass transfer
proceeds via the latter mechanism.

Although an orbital period of $\sim$3.412~days (Jones et
al. \cite{jones}) was quickly identified for \object{4U~1700-37},
extensive searches (e.g. Rubin et al. \cite{rubin} and references
therein) have failed to identify any other X-ray periodicities within
the system that might correspond to the pulse period for a possible
neutron star (although Konig \& Maisack \cite{konig} claim the
presence of a 13.81~day period in CGRO BATSE \& RXTE ASM datasets).
Given the absence of any X-ray pulsations and the unusually hard
nature of the spectrum various authors (e.g. Brown, Weingartner \&
Wijers \cite{brown}) have suggested that the compact companion could
be a low mass black hole rather than a neutron star. However, Reynolds
et al.  (\cite{reynolds}) point out that the 2-200~keV spectrum of
\object{4U~1700-37} differs from those commonly observed for black
hole candidates such as \object{Cygnus X-1}.  Given that the X-ray
spectrum of \object{4U~1700-37} is qualitatively similar to those of
accreting neutron stars they suggest that the compact object is also a
neutron star, and explain the lack of pulsations as due to either a
weak magnetic field or an alignment of the magnetic field with the
spin axis.

With a spectral type of O6.5 Iaf$^{+}$, \object{HD~153919} is the
hottest and potentially most massive mass donor of any of the the HMXB
systems. As such, determination of its fundamental parameters (radius,
temperature, mass and chemical composition) is of importance given
that these will potentially provide valuable insights into the
evolution of very massive stars and their ultimate fate.  In
particular by determining the masses of the components of HMXB
systems, limits to the progenitor masses for neutron stars and black
holes in such systems can be found\footnote{we note that Wellstein \&
Langer (\cite{wellstein}) demonstrate that such limits derived from
binary systems cannot be {\em directly} applied to single stars} while
the chemical composition of the mass donor can shed light on the
pre-supernova (SN) mass transfer mechanisms.

The paper is ordered as follows. Section 2 describes the determination
of the physical properties of \object{HD~153919}; the complete dataset
and the non-Local Thermal Equilibrium (non-LTE) code used to analyse
it. Section 3 describes the Monte-Carlo technique used to determine
the masses of both components of the binary system. In Sections 4 \& 5
we discuss the implications of our results for the evolution of hot
massive stars, limits for the progenitor masses of compact objects and
the equation of state for nuclear matter.  Finally in Section 6 we
summarize the main results of the paper.

\section{Determination of the stellar parameters for \object{HD~153919}}

Determination of the fundamental stellar parameters of
\object{HD~153919} is complicated by its high temperature and mass
loss rate, which necessitates a sophisticated non-LTE treatment.
Early attempts to determine the physical properties of
\object{HD~153919} suggested that the star might be undermassive by a
factor of 2 (e.g. Conti \cite{conti}, Hutchings \cite{hutchings})
which would imply that the terminal velocity of the stellar wind
(V$_{\infty} \sim$1700~km~s$^{-1}$; van Loon, Kaper \&
Hammerschlag-Hensberge 2001 (henceforth vL01)) is a factor of 10 in
excess of the escape velocity (V$_{\rm esc}$). Given that Howarth \&
Prinja (\cite{howarth}) show that V$_{\infty}$/V$_{\rm esc} < 4$ for O
stars (and typically $\sim$2.5) this discrepancy clearly needs to be
resolved. More recent analyses still do not resolve the issue, with
Heap \& Corcoran (\cite{heap}) suggesting a mass for
\object{HD~153919} of M$_{\ast}$=52$\pm$2~M$_{\odot}$ (thus broadly in
line with the expected mass for such a star) while Rubin et al.
(\cite{rubin}) propose M$_{\ast}$=30$^{+11}_{-7}$~M$_{\odot}$,
suggesting that the star is probably undermassive for its spectral
type.

As will be shown in Sect. 3, determination of the masses of {\em both}
components in the system is hampered by the considerable uncertainties
in the stellar radius of \object{HD~153919}. In order to address this
problem, and in the light of dramatic advances in the sophistication
of non-LTE model atmospheres we have decided to reanalyse both new and
published ultraviolet to near-infrared spectroscopic and optical to
mid-infrared photometric observations of \object{HD~153919} in order
to refine previous estimates of the stellar parameters.

\subsection{The complete dataset}

Archival and new ultraviolet to mid-infrared spectroscopic and
photometric data were used to derive a set of stellar parameters for
\object{HD~153919}. High spectral resolution UV spectroscopy was
obtained with the {\em International Ultraviolet Explorer} (IUE); the
data used and reduction procedures employed are described in Kaper et
al. (\cite{kaperb}); the details are not repeated here.  Four high S/N
and high spectral resolution (R=48,000) optical spectra
($\sim$3700-8600{\AA}) were obtained in 1999 April with the Fiber-fed
Extended Range Optical Spectrograph (FEROS) mounted on the ESO 1.52m
telescope at La Silla. All 4 spectra were wavelength calibrated and
optimally extracted to determine if significant changes in the
spectrum occured at different orbital periods. Besides line-profile
variability in the strongest ``wind'' lines and the shift in radial
velocity due to orbital motion, no evidence is found for intrinsic
variability of the photospheric spectrum. The final spectrum used for
determining the stellar parameters of \object{HD~153919} was that
taken during X-ray eclipse to further minimize the effects of any
perturbation of the wind by the presence of the compact companion (see
Sect.  2.2). Near-infrared spectra between 1-2.2$\mu$m and optical to
near-infrared photometry were taken from Bohannan \& Crowther
(\cite{bohannan}) and mid-IR photometry (6.8$\mu$m=669~mJy,
11.5$\mu$m=244~mJy) from Kaper et al. (\cite{kaperc}); see respective
papers for the particular reduction strategies employed in each case.

\subsection{Spectral analysis}

\begin{figure*}
\caption{Plots of selected regions of the optical spectrum of
\object{HD~153919} (solid line) and best fit model (dotted line);
parameters listed in Table 1.}
\resizebox{\hsize}{!}{\includegraphics[angle=-90]{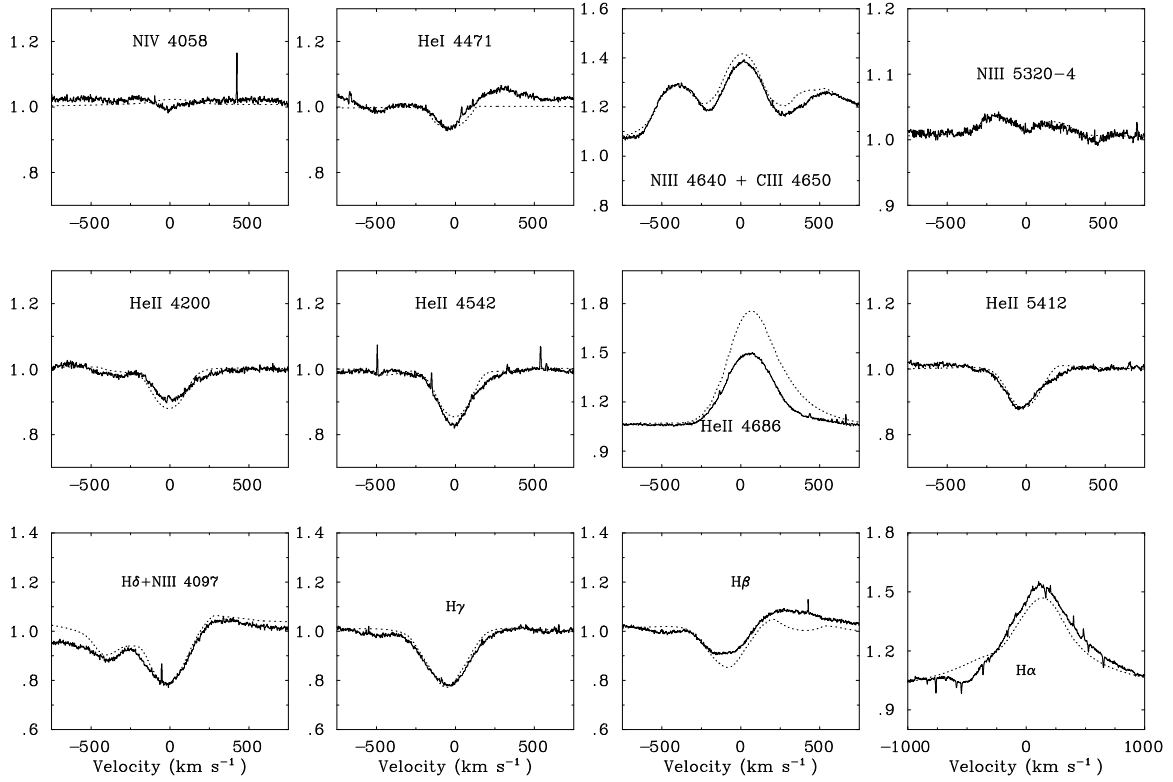}}
\label{Figure 1}
\end{figure*}      

\begin{figure}
\caption{Plot of the observed (solid line and data points) and 
theoretical (dotted line) UV - mid-IR spectral energy
distribution for \object{HD~153919}}
\resizebox{\hsize}{!}{\includegraphics[angle=-0]{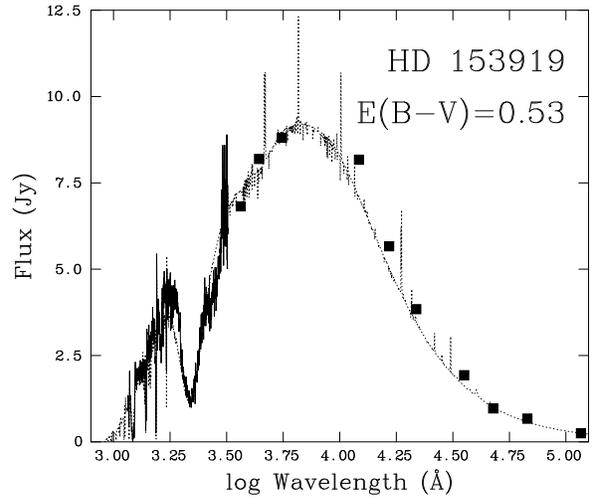}}
\label{Figure 2}
\end{figure} 

\begin{figure*}
\caption{Plots of selected regions of the UV spectrum of
\object{HD~153919} (solid line) and best fit model (dotted line);
parameters listed in Table 1. Note that our model does not take into
account the observed Raman-scattered emission lines (which are
not of a photospheric origin) in the range 1400--1700~\AA. }
\resizebox{\hsize}{!}{\includegraphics[angle=-0]{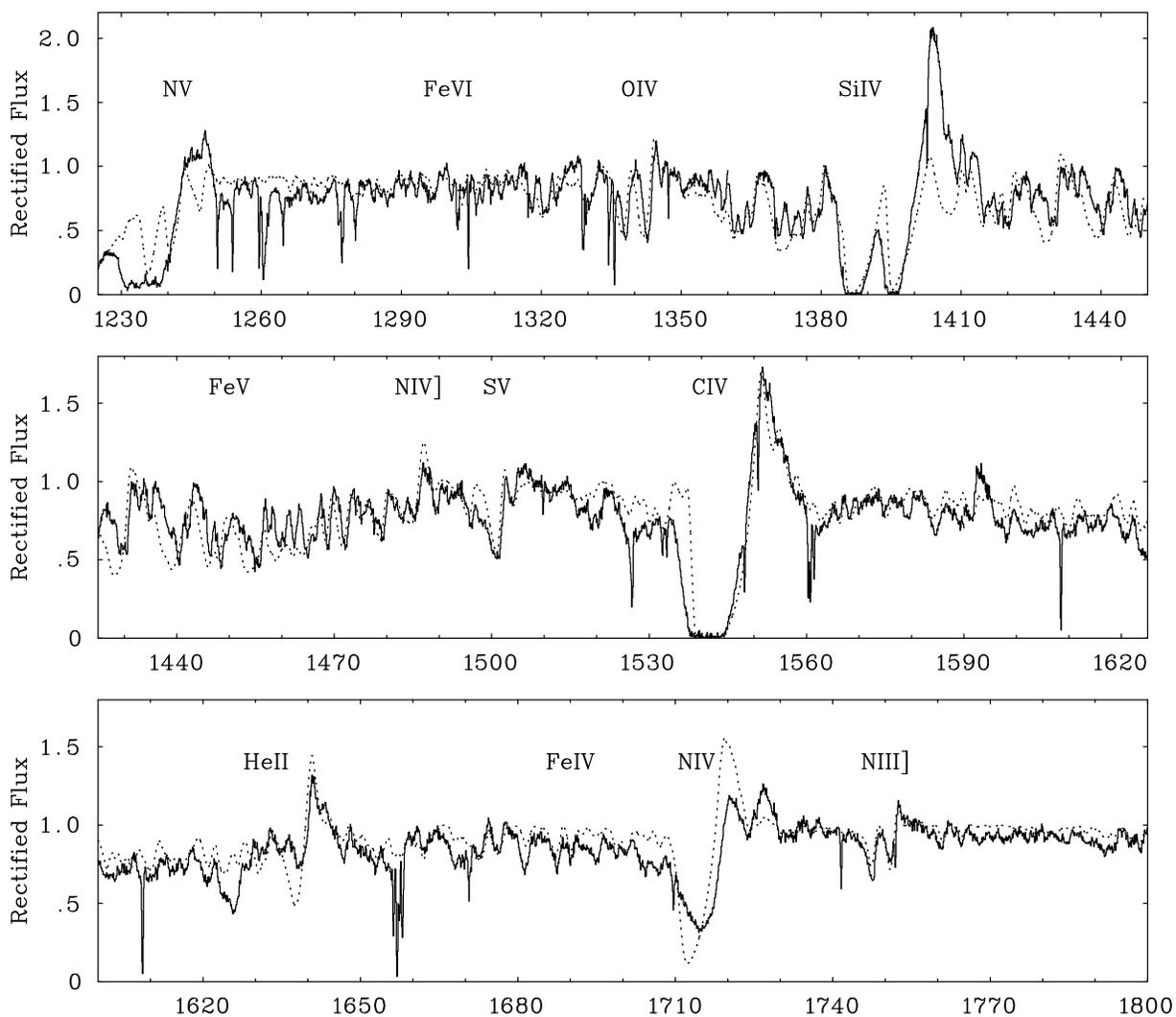}}
\label{Figure 3}
\end{figure*}  

To determine the stellar properties of \object{HD~153919} we have
utilised the non-LTE code of Hillier \& Miller (\cite{hillier}) which
solves the radiative transfer equation subject to the constraints of
statistical and radiative equilibrium, in a spherical, extended
atmosphere.  Line blanketing is incorporated directly through the use
of a super-level approach. We use a similar atomic model to that
employed by Crowther et al. (\cite{crowtherb}) in their study of early
O supergiants, including H\,{\sc i}, He\,{\sc i-ii}, C\,{\sc iii-iv},
N\,{\sc iii-v}, O\,{\sc iii-vi}, Si\,{\sc iv}, P\,{\sc iv-v}, S\,{\sc
iv-vi} and Fe\,{\sc iv-vii}.  For extreme O supergiants, line
blanketing and the strong stellar wind conspire to produce
significant differences in stellar parameters relative to the
standard plane-parallel hydrostatic results (see Crowther et
al. \cite{crowtherb} for further details).

Our procedure is as follows. We adjust the stellar
temperature\footnote{Defined, as is usual for an extended atmosphere,
as the effective temperature corresponding to the radius at a
Rosseland optical depth of 20} and mass-loss rate of an individual
model until the `photospheric' He\,{\sc ii} $\lambda$4542 and He\,{\sc
i} $\lambda$4471 lines are reproduced.  Simultaneously, we vary the
total mass-loss rate until H$\alpha$ is also matched. The exponent of
the $\beta$-law is adjusted until the shape of H$\alpha$ is well
reproduced -- for \object{HD~153919} we obtain $\beta\sim$1.3. The
input atmospheric structure, connecting the spherically extended
hydrostatic layers to the $\beta$-law wind is achieved via a
parameterized scale height, $h$ (see Hillier et al. \cite{hillierc}
for details), for which $h$=0.001 yields a reasonable match to
He\,{\sc i} and Balmer line wings, consistent with $\log
g$=3.45--3.55. We adopt a terminal wind velocity of $v_{\infty}$=1750
km\,s$^{-1}$ (vL01; Howarth et al. \cite{howarth}).

The formal solution of the radiative transfer equation yielding the
final emergent spectrum is computed separately, and includes standard
Stark broadening tables for H\,{\sc i}, He\,{\sc i-ii}. Except where
noted, these calculations assume a microturbulent velocity $v_{\rm
turb}$=10 km\,s$^{-1}$.  Hillier et al. (\cite{hillierc}) also
discuss the effect of varying $v_{\rm turb}$ in O star
models. Additionally, we find good agreement with observations using
$v \sin i$=150 km\,s$^{-1}$ (Howarth et al. \cite{howarth} derived 120
km\,s$^{-1}$).

It is extremely difficult to determine accurate He/H abundances in O
supergiants as discussed by Hillier et
al. (\cite{hillierc}). Consequently, we adopt He/H=0.2 by number,
whilst C and N abundances are varied until diagnostic optical line
profiles are reproduced. In Fig. 1 we present selected optical line
profile fits to FEROS observations of HD\,153919. Overall, agreement
for $T_{\rm eff}$=35kK is very good, with the exception of He\,{\sc
ii} $\lambda$4686. He\,{\sc i} $\lambda$4471 provides our main
temperature constraint since other blue optical He\,{\sc i} lines are
weak or absent. Alternatively, we considered using He\,{\sc i}
$\lambda$5876 (or $\lambda$10830) together with the He\,{\sc ii}
$\lambda$4686 line. However, this method (followed by Crowther \&
Bohannan \cite{crowthera}) yields significantly ($\sim$4kK) lower
stellar temperatures, and suffers from inconsistencies involving the
ionization balance of UV/optical metal lines. Therefore, we have
greater confidence in our adopted diagnostics, which do not suffer
from such problems.

From spectral energy distribution fits to IUE spectrophotometry and
Johnson photometry (Fig. 2), we derive $E_{\rm B-V}$ = 0.53$\pm$0.02
mag. Alternatively, using the intrinsic colour of $(B-V)_{0}=-$0.30
from the early O supergiant calibration of Schmidt-Kaler (\cite{sk}),
we derive $E_{\rm B-V}$=0.55 from Johnson photometry of HD\,153919
($V$=6.54 and $B-V$=0.25, Bolton \& Herbst \cite{bolton}).
Consequently, we adopt $E_{\rm B-V}$=0.54$\pm$0.02 for the remainder
of this work. We adopt a distance modulus of 11.4 mag to \object{HD~153919} 
(Ankay et al. \cite{ankay}), implying $M_{\rm V}=-$6.53 mag.  Our derived
temperature provides a bolometric correction of $-$3.3 mag, therefore
we obtain $\log (L/L_{\odot}$)=5.82 for HD\,153919, and thus $R$=21.9
R$_{\odot}$. The derived mass-loss rate is 9.5$\times 10^{-6}$
M$_{\odot}$ yr$^{-1}$, assuming that the H$\alpha$ line-forming region
is not clumped. Moderate clumping would reduce this value by a factor
of $\sim$2.  Derived parameters are listed in Table 1, and are in
reasonable agreement with those derived by vL01 through the analysis
of ultraviolet resonance lines based on the Sobolev with Exact
Integration (SEI) method.

Our primary nitrogen abundance diagnostics are N\,{\sc iii}
$\lambda$4634--41 and $\lambda$4097, which together imply
$\epsilon_{\rm N} = 9 \epsilon_{{\rm N},\odot}$. With this value, the
very weak N\,{\sc iii} $\lambda$5320--4 feature is well matched, but
other lines in the vicinity of He\,{\sc ii} $\lambda$4542 are somewhat
too strong (likely due to an incomplete treatment of N\,{\sc iii}
quartet states in our models).  Carbon is somewhat more difficult to
constrain, with C\,{\sc iii} $\lambda$4647--51 well matched for
$\epsilon_{\rm C} = 1.0 \epsilon_{{\rm C},\odot}$.  C\,{\sc iv}
$\lambda$5801--12 is well reproduced with this value, whilst C\,{\sc
iii} $\lambda$5696 is too weak, implying a yet higher abundance.  As
discussed elsewhere (e.g. Hillier et al. \cite{hillier}), oxygen is
exceedingly difficult to constrain in mid-O supergiants due to lack of
suitable optical diagnostics.  The high nitrogen overenrichment is not
easily explained via single star evolution, unless carbon (and to a
lesser degree oxygen) is very depleted via the CN (or ON) cycle.
Crowther et al. (\cite{crowtherb}) discuss similar problems for
Magellanic Cloud O supergiants.

Turning to UV comparisons, we show rectified high resolution IUE
spectroscopy of \object{HD~153919} (phase 0.15) in Fig. 3 together
with synthetic spectra. Overall, agreement for He\,{\sc ii}
$\lambda$1640, C\,{\sc iv} $\lambda$1550 and N\,{\sc iv} $\lambda$1718
is reasonable, with predicted Si\,{\sc iv} $\lambda$1393-1402 emission
too weak adopting $\epsilon_{\rm Si} = 1.0 \epsilon_{{\rm
Si},\odot}$. Since X-rays are not explicitly considered in this study,
the shocked UV N\,{\sc v} $\lambda$1238--42 resonance doublet is
predicted to be too weak.  The sole prominent oxygen feature present
in the UV (or optical) region is O\,{\sc iv} $\lambda$1338--43, which
is reasonably well matched with $\epsilon_{\rm O} = 0.5 \epsilon_{{\rm
O}, \odot}$, although we do not claim that this represents an accurate
constraint.

Additional, powerful evidence in favour of our derived temperature is
the good match between the synthetic Fe\,{\sc iv-v} spectrum and
observations, again with $\epsilon_{\rm Fe} = 1.0 \epsilon_{{\rm Fe},
\odot}$).  Fig. 2 shows good agreement with the dominant Fe\,{\sc v}
`forest' observed between $\lambda$1300--1600 in HD\,153919, plus the
weaker Fe\,{\sc iv} in the $\lambda$1500--1800 region. Fe\,{\sc vi} is
not strongly predicted nor observed in the $\lambda$1200--1400 region
(see Crowther et al.  \cite{crowtherb} for further details).

While it is possible to explain the nitrogen enrichment in terms of
rotational mixing it is impossible to produce carbon enrichment via this
mechanism. Any carbon produced in the helium burning layers of the
star has to pass through the hydrogen burning layers before reaching
the surface where it will be converted to nitrogen.  Therefore, any
excess carbon in \object{HD~153919} is therefore likely to result from
mass transfer from the more evolved binary component prior to SN -
this will be returned to in Sect. 4.

Given the presence of a compact companion for \object{HD~153919}, it
is reasonable to ask whether the assumption of spherical geometry is
justified - does the X-ray flux lead to significant departures from
spherical symmetry for the ionisation of the wind (which in turn could
lead to modifications in the line driving force)? Hatchett \& McCray
(\cite{hatchett}) suggest that the X-ray emission will lead to a
reduction of moderately ionised atoms in the wind (such as Si\,{\sc
iv} and C\,{\sc iv}). Given that the ionised zone will move with the
compact object we might expect to see orbital modulation in some of
the wind UV resonance lines as the ionised zone passes in front and
behind the stellar disc (or from the presence of a photo-ionization
wake in the system, cf.\ Kaper et al. \cite{kaper94}).  However, there
is no convincing evidence for orbital modulation in the UV resonance
line due to the Hatchett-McCray effect (e.g. Kaper et al. \cite{kapera},
\cite{kaperb}); the small changes in line profiles with orbital
phase are instead most likely due to Raman scattering of EUV photons
generated by the X-ray source (Kaper et al. \cite{kapera},
\cite{kaperb}). Additionally, the modeling was performed on the
spectrum obtained during the X-ray eclipse to further minimise any
possible effects of irradiation on the stellar wind (cf.\ Sect. 2.1).

Using a modified 2-dimensional Sobolev Exact Integration (SEI) code
vL01 analyse the UV line variability and confim that any Str\"{o}mgren
sphere caused by the presence of the X-ray source is rather small,
and will have a negligible effect on the ionization structure and line
driving of the wind (since the wind is dense and the ionizing flux
low). Equally, the Str\"{o}mgren zone does not extend to the surface of
the star and so should not lead to a significant degree of X-ray
heating of the stellar surface.

Phase resolved continuum observations (e.g. vL01,
Hammerschlag-Hensberge \& Zuiderwijk \cite{HH}, van Paradijs et
al. \cite{vanP}) constrain orbital variability to $<$4~per cent in the
UV and 4-8~per cent in the optical, indicating that \object{HD~153919}
shows little departure from sphericity (possibly as a result of a
large mass ratio). Note that continuum emission from the stellar wind
is essentially negligible at wavelengths shorter than a few
microns. Therefore, the lack of significant variability cannot be
attributed to emission from the outer regions of the stellar wind
`shielding' a heavily perturbed stellar surface and/or inner wind from
view.

The photometric variability further constrains any change in stellar
temperature due to X-ray heating to less than the uncertainty in the
stellar temperature derived from our NLTE modeling. 
Therefore, we have confidence that deviations from spherical symmetry in 
\object{HD~153919} and/or the effects of X-ray irradiation 
are negligible for the purposes of spectroscopic modeling.

\begin{table}
\begin{center}
\begin{tabular}{lr}
\hline
Parameter  & Value \\
\hline
E(B-V) & 0.54$\pm$0.02 \\
T$_{eff}$ & 35,000$\pm$1000K \\
log(L$_{\ast}$/L$_{\odot}$) & 5.82$\pm$0.07 \\
R$_{\ast}$ & 21.9$^{+1.3}_{-0.5}$~R$_{\odot}$ \\
\.{M} & 9.5$\times$10$^{-6}$~M$_{\odot}$~yr$^{-1}$ \\
v$_{\infty}$ & 1750~km~s$^{-1}$ \\         
$\log g$ & 3.45--3.55 \\
\hline
\end{tabular}
\caption{Stellar parameters for \object{HD~153919} derived from the
NLTE modeling described in Sect. 2.2.}
\end{center}
\end{table}

\section{Mass determination for the system components}

Since no X-ray pulsations have been convincingly measured for
\object{4U1700-37} - and hence no determination of $a_x$sin$i$ is
possible - the orbital solution cannot be uniquely determined. However,
following Heap \& Corcoran (\cite{heap}) and Rubin et
al. (\cite{rubin}) we may estimate the mass of the companion using a
Monte Carlo method.  The mass of the companion can be calculated using
a series of equations relating eclipse and orbital parameters.
 
The Roche lobe filling factor $\Omega$ is defined by
 
\begin{equation}
R_{\ast} = \Omega R_L
\label{eqn:1}
\end{equation}
 
\noindent where $R_{\ast}$ is the radius of the O-star and $R_L$ is the
Roche lobe radius which is related to the semi-major axis of the
system $a$ and the mass ratio $q$ (= M$_{ast}$/M$_{x}$) by

\begin{equation}
\frac{R_L}{a} = A + B \log{q} + C (\log{q})^2
\label{eqn:2}
\end{equation}
 
\noindent and the coefficients $A$, $B$ and $C$ are
 
\[
A = 0.398 - 0.026 \Gamma^2 + 0.004 \Gamma^3
\]
 
\[
B=-0.264 + 0.052 \Gamma^2 - 0.015 \Gamma^3
\]
 
\[
C=-0.023 - 0.005 \Gamma^2
\]
 
\noindent where $\Gamma$ is the ratio of the rotational angular   
frequency of the companion to its orbital angular frequency (Rappaport
\& Joss \cite{rappaport}).
 
The radius of the O-star is related to the semimajor axis by the
inclination $i$ and eclipse semiangle $\theta_E$ by
 
\begin{equation}
\frac{R_{\ast}}{a} = \sqrt{ {\rm cos}^2 i + {\rm sin}^2 i {\rm cos}^2
\theta_E}
\label{eqn:3}
\end{equation}
 
\noindent while the companion mass function $f$ is given by
 
\begin{equation}
{\rm M}_x^3 {\rm sin}^3 i = f ({\rm M}_{\ast} + {\rm M}_x)^2
\label{eqn:4}
\end{equation}
 
\noindent and $f$ is related to orbital parameters by
 
\begin{equation}
f = 1.038 \times 10^{-7} K_{\ast} P (1-e^2)^{3/2}
\label{eqn:5}
\end{equation}
 
\noindent where $K_{\ast}$ is the radial velocity semi-amplitude in km
s$^{-1}$, $P$ the period in days and $e$ the ellipticity.  Finally by
Kepler's third law
 
\begin{equation}
a^3 = 75.19 (1+q) {\rm M}_x P^2
\label{eqn:6}
\end{equation}
 
We combine these equations in a similar way to Rubin et al. (\cite{rubin}) to
obtain
 
\begin{equation}
R_{\ast}^2 = \frac{R_L^2}{R_{La}^2} - 17.81 P^{4/3} f^{2/3} (1+q)^2 {\rm
cos}^2 \theta_E
\label{eqn:7}
\end{equation}
 
\noindent where $R_{La} = R_L/a$.  This equation can then be solved
numerically for $q$.
 
Values of various system parameters (listed in Table~\ref{tab:monte}) are
selected randomly either from a gaussian distribution (if observed) or
uniformly if constrained between certain values.  Equation~\ref{eqn:7}
is solved for $q$ which gives $a$ from Eq.~\ref{eqn:2} which then
gives M$_x$ from Eq.~\ref{eqn:6}.  Consistency can be checked by
requiring sin $i < 1$ and $i > 55$ degrees (e.g. Rubin et al. \cite{rubin}). 
 
This procedure is followed $10^6$ times to gain a distribution of
M$_x$ and M$_{\ast}$ for a considerable number of possible parameter
combinations (noting that as expected there is a very strong positive
correlation between the two masses). We find that M$_x = 2.44 \pm
0.27$~M$_{\odot}$.  As shown by the histogram of M$_x$ in Fig. 4 the
distribution is very asymmetric beyond the $1 \sigma$ limits
(determined by the 16$^{th}$ and 84$^{th}$ percentile of the
cumulative distribution function), with only 3.5 per cent of the
sample having a mass of less than 2~M$_{\odot}$, and {\em none} less
than $1.65 M_{\odot}$ (which is significantly higher than the upper
limit to the range found for binary pulsars by Thorsett \& Chakrabarty
\cite{thorsett}). We note that {\em none} of the 10$^6$ trials were
rejected from inclination constraints suggesting that the range of
stellar radii adopted for the modeling are unlikely to be
significantly in error (which, for the fixed eclipse length, would
lead to unphysical solutions for the orbital inclination).

The errors on M$_x$ are significantly smaller than previous work (e.g.
Rubin et al. \cite{rubin}) due to the far more stringent limits on
$R_{\ast}$, which constrain the orbital and eclipse parameters far
more strongly.  This is not surprising as the eclipse parameters are
used to work out the orbital parameters and the eclipse constraints
rely strongly on the O-star radius.

Fig. 5 shows the O-star mass distributions around M$_{\ast} = 58 \pm 11$
M$_{\odot}$.  Again the distribution is anti-symmetric with 32 per cent
of trials between 50-60~M$_{\odot}$, 26 per cent between
40-50~M$_{\odot}$ and only 2 per cent less than 40~M$_{\odot}$.
Therefore, the mass implied for \object{HD~153919} appears to be
consistent with both that expected from its spectral classification
and relevant evolutionary tracks (see Fig. 6), and that
suggested by its high terminal wind velocity (Sect. 2). Additionally
the $\log{g}$ determined from the He\,{\sc i} and Balmer line wings
(Sect. 2.2) indicates a {\em minimum} mass of 50~M$_{\odot}$ (and
maximum of $\sim$60~M$_{\odot}$), again fully consistent with the
results of the Monte Carlo simulation. Therefore, given the
consistency between mass estimates based on spectral type,
evolutionary tracks (when compared to the stellar temperature and
luminosity derived from modeling), surface gravity and the Monte Carlo
simulations, we have confidence that the mass of \object{HD~153919}
lies in the range 50-60~M$_{\odot}$. This resolves the problem that the
star is undermassive by a factor of $\sim$2.

However, the mass of the compact companion is more problematic given
that it is {\em significantly} in excess of the observed mass range
for NS, but apparently considerably lower than those found for BH
candidates (e.g. Fig. 7). If a minimum mass of 50~M$_{\odot}$ is
adopted for \object{HD~153919} the minimum value of M$_{x}$ that may
be obtained is 1.83~M$_{\odot}$, while for   values of M$_{o}$ between
50-60~M$_{\odot}$ only 0.17 per cent of trials
result in M$_{x} <$2~M$_{\odot}$. This will be returned to in Sect. 5.

Recent reanalysis of spectroscopic data by Hammerschlag-Hensberge et
al. (in prep.) suggests that the eccentricity of the orbit is somewhat
uncertain, and that the orbital velocity curve is equally well fit by
an orbit of eccentricity $e \sim 0.22\pm0.04$ as it is by a circular
orbit. In order to address this uncertainty we modified the above
equations for the more general case of an elliptical orbit and
repeated the simulations with $e=0.22\pm0.04$. This resulted in
significantly higher masses for both components, with
M$_{\ast}$=70$\pm$7~M$_{\odot}$ and M$_x$=2.53$\pm$0.2~M$_{\odot}$. 
Therefore, the mass of the O star in the case of an elliptical orbit
is   significantly higher than expected for
an O6.5 Iaf$^{+}$ star (only 0.002~per cent of the trials result in a
mass $\leq$50~M$_{\odot}$, and 5~per cent give a mass between
50-60~M$_{\odot}$).  Such high values for M$_{\ast}$
are inconsistent with the measured $\log{g}$ and we note that 95~per
cent of trials are rejected due to the inclination constraints,
suggesting that a low eccentricity solution is more likely.

If such extreme values for M$_{\ast}$ are adopted, the mass of the
compact object is still less than that observed for the lowest mass
black hole candidate known ($\sim$4.4~M$_{\odot}$; Sect. 5) and
remains significantly greater than any known neutron star. Indeed, the
lowest mass estimates for both components were derived in the case of
a circular orbit; therefore the value of
M$_{x}$=2.44$\pm$0.27~M$_{\odot}$ represents a lower limit for the
mass of the compact object\footnote{Values of M$_x < $2 M$_{\odot}$
are only obtained when the radial velocity semi-amplitude, K$_0$, is
$>$2 $\sigma$ below the observed value, to obtain M$_x$ $<$ 1.6
M$_{\odot}$ would require K$_0$ to be wrong by several sigma.}, and we
suggest that these results favour a low eccentricity solution for the
orbit (we note that the orbital eccentricity of \object{Vela X-1} is
overestimated from optical observations when compared to the value
derived from timing analysis, cf.\ Barziv et al. \cite{barziv}).

\begin{figure}
\resizebox{\hsize}{!}{\includegraphics[angle=-90]{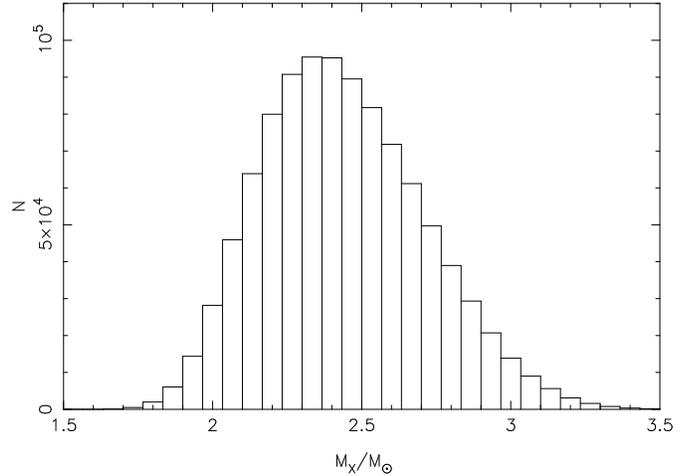}}
\caption{Histogram of the results of the Monte Carlo simulations for
the mass of the compact object in \object{4U1700-37}. The results
indicate a mass in the range of 2.44$\pm$0.27~$R_{\odot}$ with only
3.5~per cent of simulations indicating masses of less than
2~M$_{\odot}$, and {\em none} $<$1.65~M$_{\odot}$.  }
\label{Figure 4}
\end{figure}    

\begin{figure}
\resizebox{\hsize}{!}{\includegraphics[angle=-90]{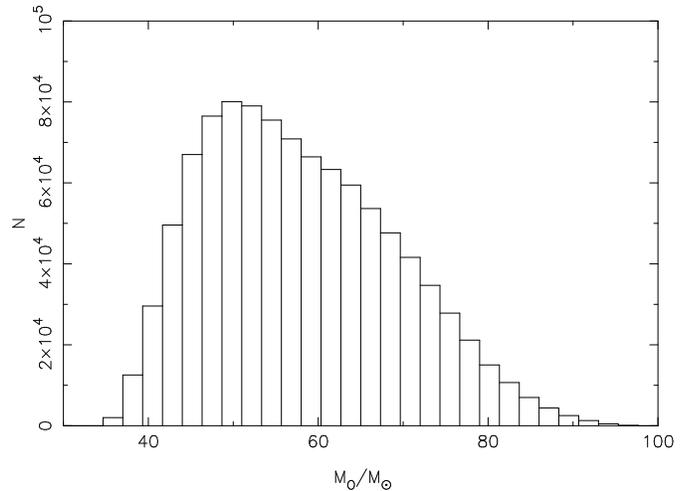}}
\caption{Histogram of the results of the Monte Carlo simulations for
 the mass of the O6.5Iaf+ primary \object{HD~153919} - the results
 indicate a mass in the range of 58$\pm$11 $R_{\odot}$ consistent with
 evolutionary predictions and the mass estimated from our
 determination of log $g$=3.45-3.55 (Sect. 2.2).  }
\label{Figure 5}
\end{figure}    

\begin{table}
\begin{center}
\begin{tabular}{lr}
\hline
Parameter  & Value \\
\hline
$R_{\ast}$      & 21.4 - 23.2 $R_{\odot}$ \\
$\Gamma$   & 0.5 - 1.0$^a$ \\
$\Omega$   & 0.8 - 1.0$^a$ \\
$e$        & $0.0^b$ \\
$\theta_E$ & $28.6 \pm 2.1$ degrees$^a$\\
$P$        & $3.411581 \pm 2.7 \times 10^{-5}$ days$^a$ \\
$K_{\ast}$      & $20.6 \pm 1.0$ km s$^{-1}$$^b$ \\
\hline
\end{tabular}
\caption{Physical parameters of 4U 1700-37.  Those with a $\pm$ have a
gaussian error distribution while those without are assumed to have a
uniform distribution.  $^a$ Rubin et al (\cite{rubin}),
$^b$Hammerschlag-Hensberge et al. (in prep.)}
\label{tab:monte}
\end{center}
\end{table}

\section{Evolutionary history of \object{4U1700-37}}

\begin{figure*}
\resizebox{\hsize}{!}{\includegraphics[angle=-90]{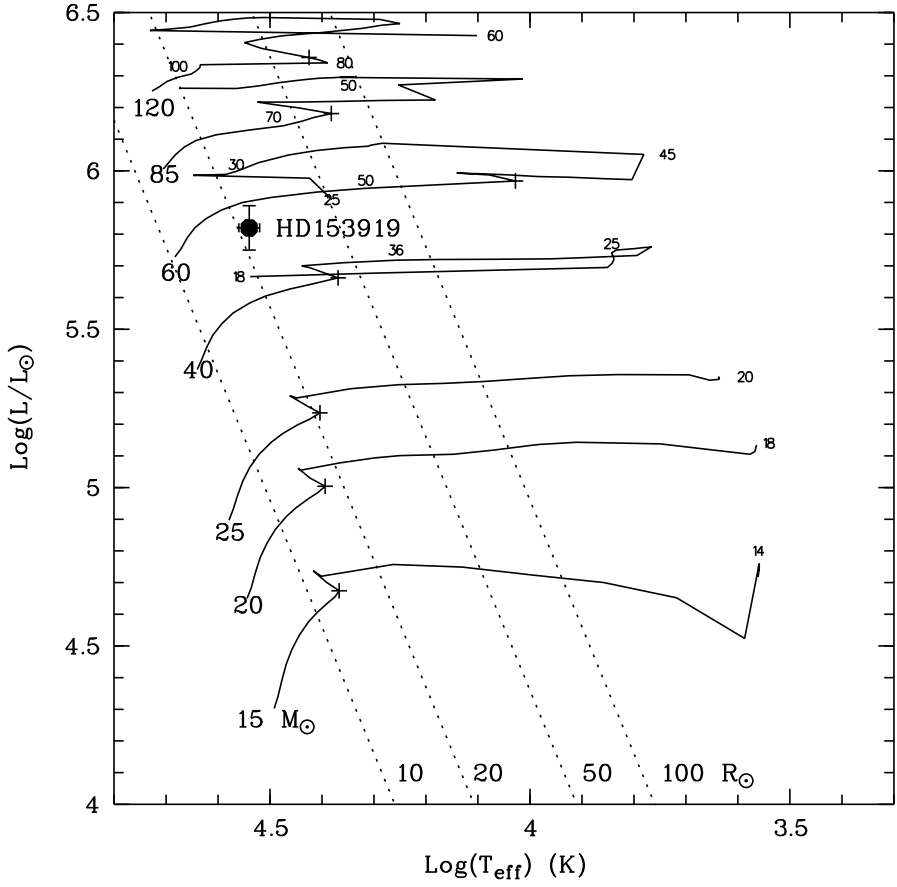}}
\caption{The location of \object{HD~153919} in the Hertzsprung-Russell
diagram. The evolutionary tracks of Lejeune \& Schaerer
(\cite{lejeune}) for stars of a given initial mass are indicated (up
to the phase of core-helium burning). The numbers along the tracks
show the decrease in M$_{\ast}$ with time due to wind losses. The plus
signs indicate the end of core-hydrogen burning.  The diagonal dashed
lines are lines of constant radius.}
\label{Figure 6}
\end{figure*}    

The stellar parameters for both primary and compact companion
determined via non-LTE modeling and Monte-Carlo simulation raise many
important questions regarding the evolution of single and binary
massive O stars and their ultimate post-SN fate. However, such
questions are complicated by the uncertain evolution of massive stars
after leaving the main sequence and the role that binarity and
associated - possibly non-conservative - mass transfer plays in
modifying this; for instance the time at which the hydrogen rich outer
layers are lost exposing the helium core plays a critical role in
determining the final pre-SN mass of the star (e.g. Brown et
al. \cite{brownb} and references therein).

Heap \& Corcoran (\cite{heap}) propose an initial
80~M$_{\odot}+$~40~M$_{\odot}$ binary system with subsequent evolution
via case B mass transfer\footnote{We adopt the nomenclature used in
Wellstein \& Langer (\cite{wellstein} and references therein), with
Case A, B \& C evolution corresponding to mass transfer during core
hydrogen burning, after core hydrogen burning but before core helium
exhaustion, and after core helium burning, respectively.}  after
2.6$\times$10$^6$~yrs, proceeding for 10$^4$~yrs (via Roche-lobe
overflow; RLOF).  After the mass transfer the initially more massive
star has lost enough material due to the combination of wind driven
mass loss and the brief period of {\em non-conservative} RLOF to
become a WR star, which subsequently explodes as a SN.

Based on their identification of the Sco OB1 association as the
birthplace of \object{HD~153919}, Ankay et al. (\cite{ankay}) propose
a lower initial mass of the SN progenitor
($\geq$30$^{+30}_{-10}$~M$_{\odot}$) based on the turnoff mass for the
proposed 6$\pm$2~Myr age of \object{Sco~OB1} {\em at the time of the
supernova}. Assuming {\em conservative} Case B mass transfer they
derive an initial mass of at least 25~M$_{\odot}$ and suggest the
short orbital period Wolf-Rayet (WR) binary \object{CQ
Cep}/\object{HD~214419} as a possible example of the progenitor
system. However they note that the assumption of conservative mass
loss might be incorrect and highlight the non-conservative scenario of
Wellstein \& Langer (\cite{wellstein}). Such a scenario is attractive
since the loss of substantial quantities of mass and angular momentum
naturally lead to short period binaries (assuming both components do
not merge). However, common envelope evolution is poorly understood
and therefore somewhat limits our ability to quantitatively
reconstruct the pre-SN evolution of the binary.

Despite these uncertainties we can address the general evolution of
the binary in some detail. Our present mass estimate for
\object{HD~153919} of M$_{\ast} = 58^{+11}_{-11}$~M$_{\odot}$ suggests
a mass for the SN progenitor of the order of $\ga$60~M$_{\odot}$, at
the upper range of that proposed by Ankay et
al. (\cite{ankay})\footnote{We note however, that the difference in
the life time of a 60~M$_{\odot}$ star (3.8~Myr) and - for example - a
120~M$_{\odot}$ star (3.0~Myr) is smaller (0.8~Myr) than the
uncertainty in the life time of a 60~M$_{\odot}$ star (1~Myr),
suggesting that higher progenitor masses {\em may} be possible.}.  The
short orbital period of \object{HD~153919}/\object{4U1700-37} favours
non conservative evolution probably via case B mass transfer.  While
mass loss via the stellar wind of the SN-precursor during the WR phase
will lead to a widening of the orbit, a favourable SN kick may
overcome these problems.

The alternative Case C evolution appears unlikely given that the SN
will occur several thousand years after the end of mass
transfer/loss. This time period appears unrealistically short given
that this will not allow sufficient time for the helium mantle to be
removed to expose the C/O core (i.e. the star will not pass though a
WC phase). Therefore the SN progenitor would have a large mass at the
point of SN; for a SN progenitor with an intial mass of 60~M$_{\odot}$
we might expect the mass to be of the order of 30~M$_{\odot}$ (the
maxium helium core mass). This implies the loss of a very large
quantity of material in the SN, contradicting the estimate of Ankay et
al. (\cite{ankay}) that only $\sim$9~M$_{\odot}$ was lost in the SN
(and we might also expect such a scenario to lead to a very high mass
compact object, cf.\ Brown et al. \cite{brownb}).

The high mass implied for the SN precursor suggests that such a star
would be likely to evolve through a Luminous Blue Variable (LBV) phase
rather than a Red Supergiant (RSG) phase on its way to becoming a WR
star (stars with initial masses $\ga$40~M$_{\odot}$ likely avoid the
RSG phase). Such an evolutionary path is likely to prevent mass
transfer {\em onto} \object{HD~153919} via RLOF (Wellstein \& Langer
\cite{wellstein}). Significant accretion of material by
\object{HD~153919} via RLOF seems implausible in any case, as this
would lead to a large orbital separation and period (Wellstein \&
Langer \cite{wellstein}).  Instead, the formation and subsequent
ejection of a common envelope ({\em despite} the SN precursor avoiding
the RSG phase) and simultaneous reduction in orbital period and binary
separation is suggested. Such a scenario therefore implies that the
present day mass of \object{HD~153919} forms a lower limit to the mass
of the SN precursor, subject to the possible accretion of a small
quantity of material directly from the wind of the SN precursor (see
below).

After the ejection of the outer layers of the SN precursor we are left
with a short orbital period WR+O star binary. Support for this
scenario is provided by the possible overabundance of carbon (or
rather the lack of significant C depletion as might be expected for
CNO processed material; Sect 2.2) in \object{HD~153919}. The carbon
rich material must have originated in the SN precursor during a carbon
rich WC stage, independently suggesting a rather high initial mass for
the SN precursor. Subsequent mass transfer would then have to occur
via direct wind fed accretion, with the wind of the WC star impacting
directly on the surface of the O star. Despite the high mass loss rate
of the O star (\.{M}=9.5$\times$10$^{-6}$~M$_{\odot}$yr$^{-1}$; Table
2) the possibility of such accretion is suggested by hydrodynamical
simulations of colliding wind binaries (Gayley, Owocki \& Cranmer
\cite{gayley}; Dessart, Petrovic \& Langer, in prep.).

We may exclude the overabundance in nitrogen originating via direct
wind accretion during the WN phase of the SN precursor.  For a
nitrogen overabundance in HD~153919 of $\sim$9, the nitrogen mass
fraction, X$_N$=0.01, is the same value as is found in the winds of WN
stars which implies that \object{HD~153919} would have to accrete 18
M$_{\odot}$ from the SN precursor during the WN phase to produce such
an overabundance. Given that this is unreasonably high, we suggest
that the excess nitrogen probably originated from internal, rotational
mixing (nitrogen overabundances are not unusual for O stars).

Several short period WR+O star binaries are known and could provide
analogues to the precursor of the present binary configuration.  At
present \object{CQ Cep} does not fit particularly well
(M$_{WN}$=21~M$_{\odot}$, M$_{O9}$=26~M$_{\odot}$, P=1.6~days) - in a
few 10$^5$~yrs when the WN star has lost mass and has evolved into a
lower mass WC star (and the period has lengthened) it may provide a
better model for the system.  However, of the known WC+O binaries,
\object{HD~63099} (M$_{WC}$=9~M$_{\odot}$, M$_{O7}$=32~M$_{\odot}$,
P=14~days) could evolve into a \object{HD~153919}/\object{4U1700-37}
like binary if the SN kick is favourable.  Other known WR+O star
binaries that could form similar systems are \object{HD~152270}
(M$_{WC}$=11M$_{\odot}$, M$_{O5-8}$=29~M$_{\odot}$, P=8.9~days) and
\object{HD~97152} (M$_{WC}$=14~M$_{\odot}$, M$_{O7}$=23~M$_{\odot}$,
P=7.9~days); in all three cases the present mass of the WC star is
$\ga$9~M$_{\odot}$ as suggested for the SN precursor by Ankay et
al. (\cite{ankay}) on the basis of the present mass of the compact
companion and the current space velocity of the system. Therefore, of
the six WC+O binaries with known masses (van der Hucht \cite{hucht})
three are found to have parameters consistent with the presumed pre-SN
stage of \object{4U~1700-37}.

The results of the Monte-Carlo simulations suggest that the SN formed
a compact object (see Sect. 5) with a mass in the range
2.44$\pm$0.27~M$_{\odot}$. As massive binary models show that Case~B
primary stars of larger initial mass evolve to a larger final mass,
and that Case~A primaries end up less massive than Case~B's (cf.
Wellstein \& Langer \cite{wellstein}; Wellstein, Langer \& Braun
\cite{wellsteinb}) we can conclude that the remnants of all Case~A/B
primaries initially less massive than about 60 M$_{\odot}$ are less
massive than about 2.5 M$_{\odot}$.

Theoretical predictions suggest that a 60 M$_{\odot}$ star in a close
binary system is capable of producing either a low or a high mass
compact object depending sensitively on the wind mass loss rate
adopted for such a star during its WR phase; a variation of only a
factor of three in the WR mass loss rate leading to compact object
masses in between 1.2 and 10~M$_{\odot}$ (Fryer et al. \cite{fryer}). This
result shows that given the present uncertainties in WR mass loss
rates, the relatively low mass found for the compact star in 4U1700-37
is not in conflict with the evolutionary scenario proposed above (and
also argues for a relatively high WR mass-loss rate).

Therefore, given the low mass of the compact companion in
\object{4U1700-37}, it seems to be difficult to explain any of the
(high mass) galactic black hole binaries as being produced through the
Case A/B channel (cf., Portegies Zwart, Verbunt \& Ergma \cite{zwart})
except for those with the most massive SN-progenitors (Brown et
al. \cite{brownb}). Indeed, evolutionary scenarios invoking Case C
mass transfer (Brown, Lee, \& Bethe \cite{browna}) seem to be required
to explain the high-mass black holes in low-mass X-ray binaries.

\section{The compact companion}

\begin{figure}
\resizebox{\hsize}{!}{\includegraphics{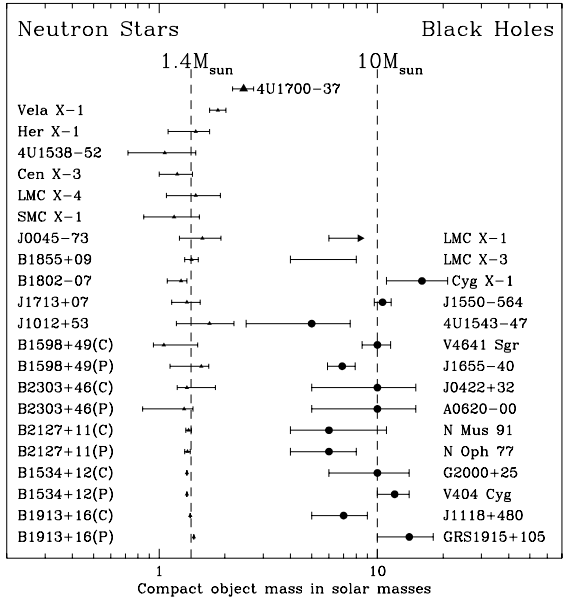}}
\caption{Mass distribution for neutron stars and black holes (after
Charles \cite{charles}).  Neutron star masses are from Ash et al. (\cite{ash}),
 Barziv et al. (\cite{barziv}), van Kerkwijk et al. (\cite{van}) and Thorsett \&
Chakrabarty (\cite{thorsett}). Black hole masses provided by Charles
(priv. comm.). Error bars for systems other than \object{4U1700-37}
are 1 sigma errors; see Sect. 3 for a discussion of the errors
associated with the mass of \object{4U1700-37} but note that the
probability of M$<$2M$_{\odot}$ is less than 3.5~per cent, and no
trials results in M$<$1.65M$_{\odot}$.}
\label{Figure 7}
\end{figure}

At present mass determinations exist for 36 compact objects, of which
21 are neutron stars and the remainder black hole candidates (Fig. 7).
Thorsett \& Chakrabarthy (\cite{thorsett}) found that the masses of
neutron stars are clustered in a remarkably narrow range (mean of
1.35M$_{\odot}$ and a standard deviation of
0.04~M$_{\odot}$). However, recent analysis of \object{Vela~X-1} by
Barziv et al. (\cite{barziv}) suggests that the pulsar has a mass of
1.87$^{+0.23}_{-0.17}$~M$_{\odot}$\footnote{Systematic excursions in
the radial velocity curve for this system complicate this
determination and prevent an {\em unambiguous} confirmation of the
mass estimate for the neutron star (Barziv et al. \cite{barziv})},
while Orosz \& Kuulkers (\cite{orosz}) find a mass of
1.78$\pm$0.23~M$_{\odot}$ for \object{Cygnus~X-2} (however Titarchuk
\& Shaposhnikov \cite{tit} have recently proposed
M=1.44$\pm$0.06~M$_{\odot}$ on the basis of Type-I X-ray bursts).

Based on assumptions about the origin of kilohertz quasi-periodic oscillations 
Zhang et al. (\cite{zhang}) suggest that several LMXBs may also 
contain massive  ($\sim$2M$_{\odot}$) neutron stars, resulting from the
accretion of substantial amounts of material over long (10$^8$~yrs)
periods of time. However their putative descendants, radio pulsar+white
dwarf binaries, provide no evidence for massive neutron stars ( Barziv et
al. \cite{barziv}).  The same is true for the Be/X-ray binaries, which 
have evolved from lower  mass systems than the OB-supergiant HMXBs. 
Only in the latter systems evidence has been found for massive neutron 
stars and black hole candidates.

Masses for a number of black hole candidates have also been
determined; lower limits to their masses comfortably exceed
3M$_{\odot}$. Indeed several objects appear to have masses
$\ga$10M$_{\odot}$ (e.g.  14$\pm$4M$_{\odot}$ for
\object{GRS~1915+105}; Greiner, Cuby \& McCaughrean \cite{greiner}),
with most typically $\ga$6M$_{\odot}$. At present the binary system
with the lowest mass candidate black hole is \object{Nova Vel}
(4.4~M$_{\odot}$; Fillipenko et al. \cite{fil}).

Given that the mass of the compact object in \object{4U1700-37} lies
outside the present observational range for both neutron stars and
candidate black holes, is it possible to increase/decrease the mass of
the compact object such that it is consistent with either type of
object? For the case of a circular orbit we find {\em no} solutions
consistent with M$_{x} \ga 4.4$~M$_{\odot}$. Very eccentric orbital
solutions do allow values of M$_{x}$ in this range although we note
there is no physical motivation for them. Such solutions imply values
of M$_{o}$ greatly in excess of 100~M$_{\odot}$, well above current
evolutionary predictions for the mass of an O6.5 Iaf+ star.  Stars
with masses in excess of 100~M$_{\odot}$ are instead expected to
evolve to H {\em depleted} WR stars via a H rich pseudo-WNL phase
where their powerful stellar winds mimic the spectra of more
(chemically) evolved stars of lower masses.

Such a large mass would also be inconsistent with the constraints
imposed by the surface gravity (Sect. 2.2) and the relationship
between stellar mass and the terminal wind velocity. Finally the
result would imply that even very massive stars (given that the
initial mass of the SN progenitor had to be significantly in excess of
the present mass of \object{HD~153919}) leave relatively low mass
remnants post supernova, presenting significant problems for the
origin of heavy ($>$10~M$_{\odot}$) black holes such as
e.g. \object{Cyg X-1}. Therefore we conclude that M$_x$ appears to be
inconsistent with the range of masses of {\em known} black hole
candidates.

Given the stringent lower limits derived in Sect. 3 it also appears
difficult to bring the mass of the compact object into line with the
range of masses found by Thorsett \& Chakrabarthy (\cite{thorsett}).
Inspection of the evolutionary tracks in Fig. 6 suggest that stars
with initial masses of the order of 60~M$_{\odot}$ initially evolve
redwards before returning bluewards after undergoing significant mass
loss, most likely during an LBV phase. While we note that the
behaviour of stars in this short lived phase is very uncertain it is
unlikely that we could be observing \object{HD~153919} after such an
excursion, since we would expect significant chemical enrichment - the
H rich mantle having been lost - which is not observed. Likewise, the
mass constraint imposed by the determination of the surface gravity
also appears to exclude this scenario.

Furthermore, such a low value for M$_{o}$ and M$_x$ would reintroduce
the problem of \object{HD~153919} being undermassive for its spectral
type by a factor $\ga$2. While the primaries in some HMXB systems are
found to be undermassive (e.g. \object{Cen X-3} and \object{LMC X-4};
Kaper \cite{kaper01}) this is attributed to mass loss via RLOF - wind
fed systems do not show this effect (Kaper \cite{kaper01}). Given that
\object{4U1700-37} is currently a wind fed system and is yet to evolve
into a RLOF system this could not explain such a mass discrepancy.
Indeed, given the evolutionary constraints imposed by the present
orbital period it is likely that no significant mass transfer has
occured onto \object{HD~153919} during its lifetime, and it will have
evolved as if it were an isolated star.

\subsection{Implications of an intermediate mass}

If we accept M$_x$=2.44$\pm$0.27~M$_{\odot}$ - as implied by the
simulations - the nature of the compact companion remains
uncertain. Conflicting claims as to the nature of the object have been
made on the basis of the X-ray spectrum of \object{4U1700-37}
(Sect. 1). While we cannot discriminate between the twin possibilities
of massive neutron star or low mass black hole from our present
measurements we note that consideration of the masses of {\em both}
components of the binary appear to exclude the possibility that stars
with masses of $\sim$60~M$_{\odot}$ can produce 5-10~M$_{\odot}$ black
holes via case A or B evolution.

If the compact object in \object{4U1700-37} is a black hole it
confirms Brown et al.'s prediction of the existence of `low mass black
holes' (based on their ``soft'' equation of state, Brown et
al. \cite{brown}), while if the object is a neutron star the high mass
would severely constrain the equation of state of matter at
supra-nuclear densities.  

The relationship between the mass of a neutron star and its central
density is calculated by integration of the Tolman-Oppenheimer-Volkoff
(TOV) equation (Oppenheimer \& Volkoff \cite{oppenheimer}) which is
the relativistic expression for hydrostatic equilibrium.  In order to
perform the integration it is neccesary to understand the equation of
state of the degenerate nuclear matter in the star.

Because of their non-perturbative nature, strong interactions between
nuclei are extremely difficult to calculate even under normal conditions.
However, there are several models of the internuclear potential in the 
literature which have achieved much success in modelling nuclei e.g. 
Stoks et al. (\cite{stoks}), Wiringa et al. (\cite{wiringa}) and 
Machleidt et al.  (\cite{machleidt}).  One of the most successful 
and up to date of these models has been applied to the neutron star 
equation of state by Akmal et al. (\cite{akmal}) yielding a maximum 
mass of between 2.2 and 2.4M$_{\odot}$ for a neutron star made completely 
of normal nuclear matter.

There is also the possibility of a QCD phase transition occuring in
the centre of the neutron star.  The large chemical potential has a
similar effect to a large temperature on the QCD coupling constant.
Consequently the interquark coupling can be reduced to the point where
deconfinement occurs and nuclei dissolve into quark matter.  The
presence of quark matter has the effect of softening the equation of
state which leads to a lower possible maximum mass for the neutron
star.  The energy scale at which deconfinement occurs can be
parameterised by the QCD bag constant $B$, a phenomenological
parameter representing the difference in energy density between the
vacua of hadronic and quark matter ($B\approx 120-200$ MeV fm$^{-3}$).
Inclusion of a phase transition to such a mixed state reduces the
maximum mass of the neutron star to $\sim$2~M$_{\odot}$ for $B=200$
MeV fm$^{-3}$ and $\sim$ 1.9~M$_{\odot}$ for $B=122$ MeV fm$^{-3}$
(Akmal et al. \cite{akmal})\footnote{ Hyperons such as $\Delta$,
$\Sigma$ and $\Lambda$ may also be produced at high density and it is
thought that their presence will also soften the equation of state and
reduce the overall permitted neutron star mass (Heiselberg \&
Hjorth-Jensen \cite{heiselberg}).}.

If the neutron is rotating rapidly, the TOV equation ceases to be a
valid approximation and one must drop the assumption of spherical
symmetry in the metric.  The effect of the rotation will increase the
allowed mass of the neutron star as one might expect, but it is shown
(Heiselberg \& Hjorth-Jensen \cite{heiselberg}) that even with
rotation one cannot obtain a neutron star with a mass much higher than
those listed above without using an equation of state so stiff that
the sound speed in the neutron-star interior becomes superluminal.

To summarize this subsection, state of the art models for nuclear
equations of state which include the effects of three nucleon
interactions marginally allow the existence of a 2.4~M$_{\odot}$
neutron star.  However, the existence of such a star would place
severe constraints upon the onset of new physics at high hadronic
densities.  
For instance, in the model where a quark matter core is
expected to develop, the bag constant $B$ would have to be
considerably larger than $200$ MeV fm$^{-3}$ in order for such a star
to be viable.

\section{Summary}

We have performed a sophisticated NLTE analysis on the O6.5 Iaf+ star
\object{HD~153919}, the primary of the HMXB \object{4U1700-37} and
used the results to constrain the masses of both components of the
system via a Monte Carlo simulation. Our NLTE model atmosphere
analysis leads to parameters for \object{HD~153919} of
log(L$_{\ast}$/L$_{\odot}$)=5.82$\pm$0.07, T$_{\rm
eff}$=35,000$\pm$1000~K, R$_{\ast}$=21.9$^{+1.3}_{-0.5}$~R$_{\odot}$,
\.{M}=9.5$\times$10$^{-6}$~M$_{\odot}$~yr$^{-1}$, and an overabundance
of nitrogen and possibly carbon over solar metallicities. Combined
with the short orbital period of the system this implies a common
envelope phase of pre-SN evolution - {\em despite the mass of the SN
progenitor apparently precluding a RSG phase} - leading to the
formation of a close WC+O star binary, with the carbon enrichment of
\object{HD~153919} a result of the impact of the stellar wind of the
WC star on the surface of the O star.

The Monte Carlo simulations result in masses for the O star and
compact object of M$_{\ast}$=58$\pm$11~M$_{\odot}$ and
M$_x$=2.44$\pm$0.27~ M$_{\odot}$, with none of the 10$^6$ trials
resulting in M$_x \leq$1.65~M$_{\odot}$, while only 3.5~per cent of
the trials result in M$_x \leq$2~M$_{\odot}$. Given that no
significant mass transfer via RLOF has occured this implies that the
initial mass of the SN precursor must have been
$\ga$60~M$_{\odot}$. Thus even very massive stars can effectively
`melt down' to leave rather low mass post-SN remnants. Equally, the
masses of both components imply that it is impossible for stars of
$\sim$60~M$_{\odot}$ to leave 5-10 M$_{\odot}$ remnants via Case A or
B evolution, suggesting that most high mass black holes are instead
formed via Case C mass transfer.

The mass of the compact object is found to lie in between the range of
masses observed for neutron stars and black holes. Given that M$_{\ast}$
and M$_x$ are strongly correlated, forcing consistency between M$_x$
and either type of object results in significant discrepancies between
M$_{\ast}$ and evolutionary predictions for the mass of
\object{HD~153919}. 

In order to produce consistency with the range of masses observed for
neutron stars the O star has to be significantly undermassive for its
spectral type and luminosity class.  While theoretical evolutionary
tracks for massive ($\ga$60~M$_{\odot}$) stars suggest that after a
redwards excursion the star will evolve bluewards again with a
substantially reduced mass, such a star would show significant
chemical enrichment (and H depletion) which is not observed.  Equally,
the surface gravity determined from modeling excludes such an
anomolously low mass.
 
Forcing consistency between M$_x$ and the masses of known black hole
candidates (M$_x \ga 4.4$~M$_{\odot}$) results in M$_{\ast}
\ga$100~M$_{\odot}$. Stars of such extreme masses are not expected to
go through an O supergiant phase, rather evolving into H depleted WR
stars via a H-rich pseudo WR phase, where their high mass loss rate
simulates the spectrum of a WR star. Once again, the constraint
implied by the surface gravity also appears to exclude this
possibility.

We are therefore left with the conclusion that no solution is fully
consistent with present expectations for stellar evolution and the
chemical abundances and surface gravity of \object{HD~153919}.  If the
compact object has a mass consistent with the observed range of
neutron star masses, the O star is significantly undermassive, while if
it consistent with the lower limit to black hole masses the O star is
overmassive by a similar (or larger) factor.  Finally if - as the
Monte Carlo analysis implies - the probable mass of the O star is
consistent with evolutionary predictions and the measured surface
gravity, the mass of the compact object lies in between the two
alternatives.

While our results do not allow us to distinguish between a massive
neutron star or a low mass black hole, the existence of a neutron star
of mass in the mass range $2.44\pm 0.27M_{\odot}$ would significantly
constrain the high density nuclear equation of state and provide
details about the QCD phase transition complementary to information
about the temperature induced transition which will be obtained at
RHIC and LHC.  Phenomena which might occur deep in the star such as
the appearance of hyperons or a quark matter core would be strongly
constrained as the existence of these phases might result in an
equation of state too soft to support such a high mass star.

\section{Acknowledgements}
This paper is partially based on observations collected at the
European Southern Observatory, Chile. The United Kingdom Infrared
Telescope is operated by the Joint Astronomy Centre on behalf of the
U.K. Particle Physics and Astronomy Research Council.  JSC, SPG \& CB
gratefully acknowledge PPARC funding.  LK is supported by a fellowship
of the Royal Accademy of Arts \& Sciences in the Netherlands.  MF is
supported by the FNRS and was helped by conversations with Nicolas
Borghini; we also thank John Porter, Marten van Kerkwijk and Phil Charles 
for their helpful comments.

\end{document}